# Equation for Calculation of Critical Current Density Using the Bean's Model with Self-Consistent Magnetic Units to Prevent Unit Conversion Errors

**Massimiliano Polichetti [1,*], Armando Galluzzi [1], Rohit Kumar [2] and Amit Goyal [2]**

[1] Laboratory "LAMBDA" for Analysis of Materials Behaviour in DC and AC fields –, Department of Physics "E.R. Caianiello", University of Salerno and CNR-SPIN Salerno, Via Giovanni Paolo II, 132, Fisciano, 84084, SA, Italy; mpolichetti@unisa.it (M.P.); agalluzzi@unisa.it (A. G.)
[2] Laboratory for Heteroepitaxial Growth of Functional Materials & Devices, Department of Chemical & Biological Engineering, State University of New York (SUNY) at Buffalo, Buffalo, 14260, NY, USA; rkumar34@buffalo.edu (R.K.); agoyal@buffalo.edu (A.G.)
* Correspondence: mpolichetti@unisa.it (M.P.)

**Abstract:** This study analyzes the calculation of the critical current density $J_{c,mag}$ by means of Bean's critical state model, using the equation formulated by Gyorgy et al. and other similar equations derived from it reported in the literature. While estimations of $J_{c,mag}$ using Bean's model are widely performed, improper use of different equations with different magnetic units and pre-factors leads to confusion and to significant errors in the reported values of $J_{c,mag}$. In this work, a SINGLE general equation is proposed for the calculation of $J_{c,mag}$ for a rectangular parallelepiped sample in perpendicular field using Bean's critical state model, underlying how the simple conversion of magnetic units can lead to a $J_{c,mag}$ in the desired units, without the need to introduce any other correction or use other specific equations depending on the units of $J_{c,mag}$. In this equation, the numerical pre-factor is dimensionless, independent of the unit system used. A comparison between the expression reported in the literature is done, showing how they can lead to different results depending on the used units, and that these results can be at least one order of magnitude different from the correct results obtained with the general equation proposed in this work. This resolves all ambiguities and aligns with the correct dimensional analysis, eliminates discrepancies in the calculated $J_{c,mag}$, and will avoid further propagation of errors in the literature.

## 1. Introduction

There is tremendous excitement for many large-scale applications of superconductors, and in particular, the energy generation via commercial nuclear fusion is presently of great interest [1,2]. There is significant research ongoing worldwide to improve the performance of superconducting wires based on various coated conductor technologies, and to lower the cost of these wires, and these have been previously reported. In such studies, one of the main elements in the characterization of the superconducting materials is the measurement of the critical current density $J_c$, which is the maximum electric current density that a material can carry without dissipation. This quantity represents one of the most important parameters for the power application of superconductors, and its experimental evaluation is of paramount interest. This evaluation can be performed both by using electric transport techniques and by means of magnetic measurements. In particular, given the high values of the involved electric current and the technical difficulties in performing electric transport measurements of $J_c$, very often its measurement by magnetic techniques, which contactless induce such high values of current, is preferred. However, the calculation of critical current densities, $J_{c,mag}$, based on the magnetization of a coated conductor represented as a rectangular parallelepiped sample in perpendicular magnetic field, poses a challenge in terms of expressions used and self-consistency of magnetic units used in past studies and reported in the literature. A commonly used expression to calculate the critical current density in a majority of these publications is the Bean Critical State Model [3,4]. However, since Bean did not report a specific expression to calculate $J_{c,mag}$ in the considered geometry, the actual formula, used in most prior publications, is the equation shown below (at a given field and temperature), reported by Gyorgy et al. [5] and derived on the basis of Bean's model. This formula has been used extensively in numerous publications in the literature, only a few of which are cited here [6–34]:

$$J_{c,mag}(B,T) = \frac{20 \times \Delta M}{w \times \left(1 - \frac{w}{3l}\right)} \tag{1a}$$

where the current density, $J_{c,mag}$, is typically desired in [A/cm$^2$], and the unit-volume full-width of the hysteresis loop for decreasing and increasing applied magnetic field $\Delta M = \mu(B-,T) - \mu(B+,T)/(w \times l \times t)$ is typically measured in [emu/cm$^3$] (i.e., in Gaussian units). The dimensions for the width $w$, the length $l$, and the thickness $t$ are typically measured in [cm]. The calculation of critical current density based on magnetization poses a challenge in terms of self-consistency of the magnetic units used in past studies and reported in the literature.

In their equation for the calculation of $J_{c,mag}$ in the general case of anisotropic critical currents, Gyorgy et al. [5] explicitly state that ΔM is in gauss, and when the field is applied perpendicular to one surface of dimensions $l$ by $t$ of a rectangular parallelepiped sample, it is expressed as follows in Figure 1:

$$\Delta M = \frac{J_{c1}\, t}{20}\left(1 - \frac{t}{3l}\frac{J_{c1}}{J_{c2}}\right), \tag{1}$$

where $\Delta M$ (in gauss) is the width of the hysteresis loop for increasing and decreasing $H_a$, the current densities are in A/cm$^2$ and the dimensions in cm. For $l \gg J_{c1}t/3J_{c2}$,

**Figure 1.** Expression for the calculation of $J_{c,mag}$ as reported in Gyorgy et al. paper (expression 1, page 283), in the general case for anisotropic critical current densities $J_{c1} \neq J_{c2}$.

Assuming that $J_{c1} = J_{c2} = J_c$, the expression (1) in Figure 1 becomes:

$$\Delta M = \frac{J_c t}{20}\left(1 - \frac{t}{3l}\right) \tag{1b}$$

From Figure 1, it is clear that if ΔM is expressed in ***gauss***, and the dimensions in cm, the resulting current density is in A/cm$^2$, without any other conversion of units. This can be checked also by considering the data and the specific examples reported in reference [5].

In practice, magnetometers like the SQUID Magnetometer and the PPMS (VSM and ACMS options) provide the magnetic moment in emu, and the sample dimensions are in cm. The interest is in the estimation of the current density, $J_{c,mag}$, in A/cm$^2$. The moment is converted to volume magnetization, M, by dividing with the sample volume, giving units of M to be emu/cm$^3$.

In order to use the expression reported by Gyorgy et al. [5] with the correct magnetic units, $\Delta$M in emu/cm$^3$ has to be converted to gauss. A standard table of magnetic unit conversions is shown in databases of IEEE and NIST [35,36], where (Volume) Magnetization M can be in emu/cm$^3$, and when M is multiplied by $4\pi$, the (Volume) Magnetization is in gauss (G). Then, the calculation using the equation reported in reference [5] gives $J_{c,mag}$, in A/cm$^2$ [37–41].

Following the equation for $J_{c,mag}$ reported in many works in literature, where the magnetization is expressed in emu/cm$^3$ [6–34], the second route to obtaining the magnetic critical current density in A/cm$^2$ is by converting $\Delta$M in emu/cm$^3$ to SI units of A/m with the expression 1 emu/cm$^3$ = $10^3$ A/m or 1 emu/cm$^3$ = 10 A/cm [35,36]. With this conversion of $\Delta$M to A/cm, using expression (1a), yields $J_{c,mag}$, in A/cm$^2$ [6–34].

The third route to use the expression (1a) is to have $\Delta$M in emu/cm$^3$, and assume that the number 20 has the units of A·cm$^2$/emu. Although not reported in regularly published works, this procedure can be found on some papers deposited on freely accessible online repositories of electronic preprint manuscripts [42,43].

There is also a fourth route, using the expression (1a) with $\Delta$M in emu/mm$^3$, and the sample's dimensions in mm, with some unclearness between the use of the magnetic moment or the magnetization, and the obtained $J_{c,mag}$ in A/mm$^2$ [44,45]. Probably other routes are also present in the literature, but at the moment, they have not been detected and reported in this paper yet.

In the first and second cases above, whether $\Delta$M is expressed in gauss or converted (as should be performed) from emu/cm$^3$ to A/cm, an error is obtained when using the expression (1a) by Gyorgy et al., which is the commonly cited formula in the literature to calculate $J_{c,mag}$ using Bean's modelIn fact, in the first case when $\Delta$M is converted to gauss, a higher $J_{c,mag}$ by a factor 12.56 is obtained, whereas, in the second case, the $J_{c,mag}$ results are 10 times higher. Although the third route provides the correct numerical value, it is conceptually wrong because it artificially introduces a non-dimensionless pre-factor 20 [A·cm$^2$/emu], not predicted by the theory, just to force the dimensional balance of an equation with non-coherent units (as will be shown in the following). In addition, if different magnetic and/or dimensional units are used, the pre-factor of 20 needs to have different "assumed" units, as it should be for the equation used in the fourth route above, as an example.

The scenarios cited above lead to confusion and errors in the calculated $J_{c,mag}$ using Bean's model. It is difficult to ascertain how the equation reported by Gyorgy et al. has been applied in numerous and extensive previous publications, since in most cases, the measured moment and/or volume magnetization and sample dimensions are typically not reported. However, it is likely that such errors have been made in previous studies, because it is unusual to expect the pre-factor, an integer number, to have complicated and non-intuitive units of A·cm$^2$/emu (or A·mm$^2$/emu, or A·cm$^2$/Gauss or whatever, depending on the authors), especially given that these units for the pre-factor 20 are not clearly specified in the literature.

The four cases reported above, definitely not exhaustive, are summarized in Table 1:

**Table 1.** Some of the different cases reported in the literature to calculate the critical current density $J_{c,mag}$ by starting from the magnetic moment measured in emu.

| | Equation | $J_{c,mag}(B,T) = \frac{20 \times \Delta M}{w \times \left(1 - \frac{w}{3l}\right)}$ <br> $M$ = Magnetization; $w$ = sample's width; $l$ = sample's length | | | |
|---|---|---|---|---|---|
| | Case | I | II | III | IV |
| UNITS FOR | ΔM | Gauss | emu/cm³ | emu/cm³ | emu/mm³ |
| | Sample lengths | Cm | cm | cm | mm |
| | $J_{c,mag}$ | A/cm² | A/cm² | A/cm² | A/mm² |
| | Pre-factor 20 | dimensionless | dimensionless | A·cm²/emu | dimensionless |
| | References | [37–41] | [6–34] | [42,43] | [44,45] |
| | Note | The emu/cm³ has to be converted to gauss, and the corresponding $J_{c,mag}$ results 12.56 times higher than the correct one. | The emu/cm³ has to be converted to A/cm, and the corresponding $J_{c,mag}$ results 10 times higher than the correct one. | No unit conversions are applied, and the obtained numerical value for $J_{c,mag}$ is correct, but the pre-factor 20 is forced to have dimensions, in contrast with the theory. | The equation cannot be applied with the reported units and pre-factor for the considered geometry. The obtained $J_{c,mag}$ is incorrect. |

## 2. Methods

The correct formula [46] for calculation of $J_{c,mag}$ starting from the *Magnetization* M, within Bean's critical state model, for a rectangular parallelepiped sample with isotropic critical current densities and with volume V = ($w \times b \times d$), when the magnetic field is applied perpendicular to the largest face with dimension $w$ by $b$ (with $w \leq b$), and with *self*-consistent magnetic units that should be used is:

$$J_c = \frac{2 \times \Delta\mu}{V \times w \times \left(1 - \frac{w}{3b}\right)} \quad (2)$$

If Δμ (where $\mu$ is the magnetic moment) is expressed in [A m²], the volume $V = (w\ b\ d)$ is in [m³], and the sample side lengths $w$ and $b$ are in [m], the ΔM results in $\frac{\mathrm{A}}{\mathrm{m}}$ and the $J_c$ is expressed in SI units as $\frac{\mathrm{Am^2}}{\mathrm{m^3 \times m}} = \frac{\mathrm{A}}{\mathrm{m^2}}$. Here, the number 2 is *dimensionless*. So:

$$\frac{2 \times \Delta M[\mathrm{A/m}]}{w[\mathrm{m}] \times \left(1 - \frac{w[\mathrm{m}]}{3b[\mathrm{m}]}\right)} = J_{c,mag}[\mathrm{A/m^2}] \quad (3)$$

As indicated in ref. [35,36], it is possible to convert the magnetic moment from [A m²] to emu, and the lengths from m to cm, by means of:

$$\Delta\mu[\mathrm{emu}] = (10^3 \mathrm{emu/(Am^2)})\Delta\mu[\mathrm{Am^2}]$$

$$V[\mathrm{cm^3}] = (10^6 \mathrm{cm^3/(m^3)})V[\mathrm{m^3}]$$

and $w[\mathrm{cm}] = (10^2 \mathrm{cm/(m)})w[\mathrm{m}]$.

Using these conversions, the $J_c$ in the Equation (2) can be expressed as $\frac{\mathrm{emu}}{\mathrm{cm}^3} \times \frac{1}{\mathrm{cm}} = \frac{\mathrm{emu}}{\mathrm{cm}^4}$ with a conversion factor of

$$\frac{10^3}{10^6} \times \frac{1}{10^2} = 10^{-5}$$

with respect to the Equation (2) in SI units.

But $\frac{\mathrm{emu}}{\mathrm{cm}^3} = 10^3 \frac{\mathrm{A}}{\mathrm{m}}$ and, by considering the simple final conversion from $\frac{\mathrm{A}}{\mathrm{m}}$ to $\frac{\mathrm{A}}{\mathrm{cm}}$, it is possible to write: $\frac{\mathrm{emu}}{\mathrm{cm}^3} = 10^1 \frac{\mathrm{A}}{\mathrm{cm}}$.

In this way, the unit conversion from the cgs emu system to some kind of "hybrid" cgs/SI system (or "practical CGS system") is obtained. With this conversion, the $J_c$ in Equation (2) expressed as $\frac{\mathrm{emu}}{\mathrm{cm}^3} \times \frac{1}{\mathrm{cm}}$ with a conversion factor of $10^{-5}$, can be expressed in $\frac{\mathrm{A}}{\mathrm{cm}} \times \frac{1}{\mathrm{cm}} = \frac{\mathrm{A}}{\mathrm{cm}^2}$ with a conversion factor of $10^{-5} \times 10^1 = 10^{-4}$ with respect to the equation in SI units, which is exactly what is expected if the conversion $\frac{\mathrm{A}}{\mathrm{m}^2} = 10^{-4} \frac{\mathrm{A}}{\mathrm{cm}^2}$ is applied to the final result obtained in SI units from Equation (2). So, the conversion from the results of Equation (2) expressed in a different unit system is simply performed by correctly applying the unit conversions [35,36], as for any formula in science, without the need to create different equations for different unit systems and add dimensional pre-factors to the existing equation.

So, if the magnetic moment is in emu and the sample's dimensions in cm, the described approach leads to writing Equation (2) by making the used units explicit:

$$J_{c,mag} = \frac{2 \times \Delta M[\mathrm{emu/cm^3}]}{w[\mathrm{cm}] \times \left(1 - \frac{w[\mathrm{cm}]}{3b[\mathrm{cm}]}\right)} \quad (4)$$

where it is evident that an expression having the dimensions of ($[\mathrm{emu/cm^3}]/\mathrm{cm}$) gives as a result a quantity having the dimensions of $[\mathrm{emu/cm^4}]$, definitely not practical for a current density. For this reason, this form of Equation (4) can be re-written in more practical units, by the previously indicated conversion $\frac{\mathrm{emu}}{\mathrm{cm}^3} = 10^1 \frac{\mathrm{A}}{\mathrm{cm}}$, which makes the dimensionless number 2 in Equation (4) become the dimensionless number 20 in the 2nd form of Equation (4) and, therefore, the current density to be expressed in the more practical dimensions of $[\mathrm{A/cm^2}]$:

$$\frac{20 \times \Delta M[\mathrm{A/cm}]}{w[\mathrm{cm}] \times \left(1 - \frac{w[\mathrm{cm}]}{3b[\mathrm{cm}]}\right)} = J_{c,mag}[\mathrm{A/cm^2}] \quad (5)$$

The above discussion clearly shows that Equation (2) is the correct one for the sample geometry described in the text, and it can be used independently of the fact that one has the magnetic moment in [A m²] (and dimensions in meters) or in [emu] (and dimensions in centimeters), by using the conversion of units, without adding any non-dimensionless pre-factor:

- When the magnetic moment is in [A m²] and the dimensions in meters, their use in Equation (2) leads to Equation (3) where the current density is directly in $[\mathrm{A/m^2}]$
- When the magnetic moment is in [emu] and the dimensions in centimeters, their use in Equation (2) leads to Equation (4) where the current density is in the unpractical $[\mathrm{emu/cm^4}]$. In order to have the current density expressed in more practical units, the conversion $\frac{\mathrm{emu}}{\mathrm{cm}^3} = 10^1 \frac{\mathrm{A}}{\mathrm{cm}}$ can be used, which introduces a multiplicative dimensionless term 10 that transforms the number 2 into 20, and leads to a current density expressed in $[\mathrm{A/cm^2}]$ (Equation (5)). On the contrary, the expression $J_c = \frac{20\times\Delta M}{w\times\left(1-\frac{w}{3l}\right)}$ where ΔM is stated to be in emu/cm³ and $J_c$ is resulting in A/cm² is misleading, because the dimensions on the right and left-hand side of the equation are not coherent and erroneously suggest the need for a further unit conversion.

This approach is also consistent with Poole et al. [46], where the equations to obtain the critical current density using Bean's model are reported for several geometries. These equations are obtained by starting from the general Equation (13.24) in ref. [46] for the magnetic moment of an arbitrarily shaped sample, and Poole et al. say that: "Equation (13.24) is an SI formula in which current density $j$ is in amperes per square meter, magnetic field $B$ is in tesla, and lengths are in meters. When practical units are used, whereby $j$ is measured in A/cm², magnetic field in gauss, and length in centimeters, the factor $\frac{1}{2}$ in Equation (13.24) is replaced by 1/20. To convert the formulae below for the magnetic moment to practical units simply divide by 10".

In fact, let us start from the definition of magnetic moment $\mu$, as reported in ref. [46]-page 396, Equation (13.24):

$$\mu = \frac{1}{2}\int_V [r \times j(r)]d^3r \tag{6}$$

and for simplicity in the notation let us denote the integrated quantity as $X$, so: $X = \int_V [r \times j(r)]d^3r$.

The Equation (6) is an SI equation, the pre-factor $\frac{1}{2}$ is dimensionless, and if the current density $j$ is expressed in A/m², and the lengths in meters (m), the value of $X$ results to be expressed in:

(m) × (A/m²) × (m³) = (A m⁴/m²) = A m², coherently with the dimension of the magnetic moment $\mu$ in SI [35,36].

Following what was reported in ref. [46], when practical units are used, the current density $j$ is expressed in A/cm², and the lengths in centimeters (cm) the value of $X$ results to be expressed in: (cm) × (A/cm²) × (cm³) = (A cm⁴/cm²) = A cm². In this way, the Equation (6) in practical units appears as:

$$\mu[\mathrm{emu}] = \frac{1}{2}X[\mathrm{A\ cm^2}] \tag{7}$$

But, emu = $10^{-3}$ (A m²) = 10 (A cm²) [35,36] and so Equation (7) is:

$$10\mu[\mathrm{A\ cm^2}] = \frac{1}{2}X[\mathrm{A\ cm^2}] \tag{8}$$

Or

$$\mu[\mathrm{A\ cm^2}] = \frac{1}{20}X[\mathrm{A\ cm^2}] \tag{9}$$

and for the magnetization $M = \mu/V$:

$$M = \frac{\mu[\mathrm{A\ cm^2}]}{V[\mathrm{cm^3}]} = \frac{1}{20}\frac{\int_V [r \times j(r)]d^3r}{V}\left[\frac{\mathrm{A}}{\mathrm{cm}}\right] \tag{10}$$

where the current density $j$ is expressed in A/cm², the lengths in centimeters (cm), and the pre-factor 1/20 is still dimensionless, coherently with Equation (5). Following this approach, as also reported by Poole et al. [46] to obtain in particular the Equation (13.30) at page 396, the expressions (2-5) can be calculated with the proper units and numerical pre-factors.

Since it has been analyzed how to perform the calculation of $J_c$ both in SI and in the 'hybrid' cgs-emu/SI system, it is also worth considering what happens in the emu system, where electric current is not an independent physical quantity and can be expressed in abampere (abA), also called biot (Bi). More precisely [47]:

$$\mathrm{abA} = \mathrm{Bi} = \mathrm{cm}^{1/2}\mathrm{g}^{1/2}\mathrm{s}^{-1}$$

The conversion to SI electric current unit is [47]: 1Bi = 10 A.

Remembering that the term "emu" used as a unit for the magnetic moment $\mu$ is not a unit but just indicates electromagnetic units, it is important to specify that in the emu system, the unit for the magnetic moment $\mu = \pi I a^2$ of a circular loop of radius $a$ carrying a current $I$ is [erg/Gauss], which, of course, as indicated in Goldfarb's paper [48], can be expressed in terms of Bi as:

$$[\mathrm{erg/Gauss}] = [\mathrm{Bi} \cdot \mathrm{cm}^2]$$

Therefore, the Equation (2) above can be used in emu system by indicating $\Delta\mu$ in $[\mathrm{Bi} \cdot \mathrm{cm}^2]$, the volume $V$ in $[\mathrm{cm}^3]$ and the length in $w$ in [cm], so leading to a current density expressed in $\frac{\mathrm{Bi \cdot cm^2}}{\mathrm{cm^3}} \times \frac{1}{\mathrm{cm}} = \frac{\mathrm{Bi}}{\mathrm{cm^2}}$, i.e.,

$$J_{c,mag} = \frac{2 \times \Delta M[\mathrm{Bi/cm}]}{w[\mathrm{cm}] \times \left(1 - \frac{w[\mathrm{cm}]}{3b[\mathrm{cm}]}\right)} = J_{c,mag}\left(\frac{\mathrm{Bi}}{\mathrm{cm}^2}\right) \tag{11}$$

that is definitely not practical. Since 1Bi = 10 A, the expression above can be written also as:

$$J_{c,mag} = \frac{2 \times 10 \times \Delta M[\mathrm{A/cm}]}{w[\mathrm{cm}] \times \left(1 - \frac{w[\mathrm{cm}]}{3b[\mathrm{cm}]}\right)} = \frac{20 \times \Delta M[\mathrm{A/cm}]}{w[\mathrm{cm}] \times \left(1 - \frac{w[\mathrm{cm}]}{3b[\mathrm{cm}]}\right)} = J_{c,mag}[\mathrm{A/cm}^2] \tag{12}$$

namely again the expression for $J_{c,mag}$ in practical units, where the pre-factor 20 is still dimensionless.

After proving the non-dimensionality of the pre-factor 20, it is probably easier to interpret the Equations (2-5):

– Starting from Equation (2) above in this paper, with the use of the SI units system where the magnetic moment $\mu$ is expressed in $[\mathrm{Am}^2]$ and the dimensions are in meters [m], the calculation is direct and the critical current density $J_{c,mag}$ is directly obtained in $[\mathrm{A/m}^2]$. Since many magnetometers present their output in emu (although in several models it is possible to choose the units for the output), the same Equation (2) can be used by inserting the magnetic moment $\mu$ in emu and the dimensions in centimeters [cm], leading for $J_{c,mag}$ a certain value (for practical reasons, let us call "Z" this value, simply resulting from the numbers given to the quantities in Equation (2)) in $[\mathrm{emu/cm}^4]$. If the value "Z" in $[\mathrm{emu/cm}^4]$ is multiplied by 10, because of the conversion

$$\frac{\mathrm{emu}}{\mathrm{cm}^3} = 10\ \frac{\mathrm{A}}{\mathrm{cm}}$$

the $J_{c,mag}$ results in $[\mathrm{A/cm}^2]$.

This leads to Equation (4) being converted into Equation (5), without adding any non-dimensionless pre-factors. This is also what Poole et al. [46] mean when at page 396 they write "To convert the formulae below for the magnetic moment to practical units simply divide by 10".

To conclude, no matter if one has the physical quantities expressed in SI, in practical units or in emu, the equation to be used (for the geometrical conditions already stated above) is Equation (2), and depending on the final units that are needed for the current density, one has just to use the standard conversion units [35,36,47,48], avoiding to make the error of mixing non-coherent units, and to compensate that error by artificially introducing dimensional pre-factors to force the dimensional balance of the equation. This will clarify any doubts about the application of the correct equations and units for the calculation of the critical current density from magnetic measurements, in particular, for a rectangular parallelepiped sample in perpendicular field, although the same approach can also be extended to analogous equations for different geometries [46].

## 3. Results and Discussion

In order to check the validity of the discussed approach, the consistency of the measurement units, and the absence of any non-dimensionless pre-factor in the expression for the calculation of the critical current density by Equation (2) above in any unit system, it can be helpful to verify the obtained results by making an example of the typical calculation of $J_{c,mag}$, if one has the magnetic moment and the dimensions of the sample.
Let us consider a hypothetical sample with the characteristics reported in Table 2:

**Table 2.** Characteristics of a hypothetical sample considered to make an example of calculation of $J_c$.

| Quantity | Value in Emu-Cgs | Value in SI System |
|---|---|---|
| Thickness $d$ = 200 nm | $2 \cdot 10^{-5}$ (cm) | $2 \cdot 10^{-7}$ (m) |
| Width $w$ = 4 mm | $4 \cdot 10^{-1}$ (cm) | $4 \cdot 10^{-3}$ (m) |
| Length $b$ = 4 mm | $4 \cdot 10^{-1}$ (cm) | $4 \cdot 10^{-3}$ (m) |
| Volume ($wbd$) | $32 \cdot 10^{-7}$ (cm$^3$) | $32 \cdot 10^{-13}$ (m$^3$) |
| Δ(Magnetic moment) = Δ $\mu$ | 3 emu | $3 \cdot 10^{-3}$ (A $\cdot$ m$^2$) |

Within Bean's critical state model, for a rectangular parallelepiped sample with isotropic critical current densities and with volume $w \times b \times d$, when the magnetic field is applied perpendicularly to the largest face with dimension $w$ by $b$ (with $w \leq b$), the equation to be used for $J_{c,mag}$ is Equation (2) above in this paper (or equivalently Equation (13.30) in ref. [46]).

So, by considering $\mu$ in emu and the dimensions in cm, the following calculations apply:

$$J_c = \frac{2 \times \Delta\mu}{V \times w \times \left(1 - \frac{w}{3b}\right)} = \frac{2 \times 3\ \text{emu}}{32 \cdot 10^{-7}\ (\text{cm}^3) \times 4 \cdot 10^{-1}\ (\text{cm}) \times \left(\frac{2}{3}\right)}$$

$$= \frac{6\ \text{emu}}{128 \cdot 10^{-7}\ (\text{cm}^3) \times 10^{-1}\ (\text{cm})} \times \frac{3}{2} = 0.0703 \cdot 10^8\ \frac{\text{emu}}{\text{cm}^4}$$

Using the conversion $\frac{\text{emu}}{\text{cm}^3} = 10\ \frac{\text{A}}{\text{cm}}$, this can be written as:

$$J_c = 0.0703 \cdot 10^8\ \frac{\text{emu}}{\text{cm}^3} \frac{1}{\text{cm}} = 0.0703 \cdot 10^8 \cdot 10^1\ \frac{\text{A}}{\text{cm}} \frac{1}{\text{cm}} =$$

$$= 0.0703 \cdot 10^9\ \frac{\text{A}}{\text{cm}^2} = 7.03 \cdot 10^7\ \frac{\text{A}}{\text{cm}^2}$$

and, since $1\ \frac{\text{A}}{\text{cm}^2} = 10^4\ \frac{\text{A}}{\text{m}^2}$

$$J_c = 7.03 \cdot 10^7 \, \frac{\mathrm{A}}{\mathrm{cm}^2} = 7.03 \cdot 10^7 \cdot 10^4 \, \frac{\mathrm{A}}{\mathrm{m}^2} = 7.03 \cdot 10^{11} \, \frac{\mathrm{A}}{\mathrm{m}^2}$$

On the other hand, by considering $\mu$ in $(\mathrm{A} \cdot \mathrm{m}^2)$ and the dimensions in m, the following calculations apply:

$$J_c = \frac{2 \times \Delta\mu}{V \times w \times \left(1 - \frac{w}{3b}\right)} = \frac{2 \times 3 \cdot 10^{-3}\ (\mathrm{A} \cdot \mathrm{m}^2)}{32 \cdot 10^{-13}\ (\mathrm{m}^3) \times 4 \cdot 10^{-3}\ (\mathrm{m}) \times \left(\frac{2}{3}\right)} =$$

$$= \frac{6 \cdot 10^{-3}\ (\mathrm{A} \cdot \mathrm{m}^2)}{128 \cdot 10^{-13}\ (\mathrm{m}^3) \times 10^{-3}\ (\mathrm{m})} \times \frac{3}{2} =$$

$$= 0.0703 \cdot 10^{13} \, \frac{\mathrm{A}}{\mathrm{m}^2} = 7.03 \cdot 10^{11} \, \frac{\mathrm{A}}{\mathrm{m}^2}$$

as obtained above, and since $1 \, \frac{\mathrm{A}}{\mathrm{m}^2} = 10^{-4} \, \frac{\mathrm{A}}{\mathrm{cm}^2}$

$$J_c = 7.03 \cdot 10^{11} \, \frac{\mathrm{A}}{\mathrm{m}^2} = 7.03 \cdot 10^{11} \cdot 10^{-4} \, \frac{\mathrm{A}}{\mathrm{cm}^2} = 7.03 \cdot 10^7 \, \frac{\mathrm{A}}{\mathrm{cm}^2}$$

as obtained above, with self-consistency and no ambiguity with respect to the magnetic units on the left and right side of Equation (2), without modifying the equation depending on the used unit system, and without artificially introducing any dimensional pre-factors to force the dimensional balance of the equation.

A comparison with the results obtained in the four cases reported above in this paper is shown in Table 3:

**Table 3.** Comparison of the results obtained with the use of the proposed Equation (2) and those obtained with the four cases illustrated above and reported in the literature.

| Equation | | | | | $J_{c,mag} = \frac{2 \times \Delta M}{w \times \left(1 - \frac{w}{3b}\right)}$ <br> *w* = sample's width; <br> *b* = sample's length | |
|---|---|---|---|---|---|---|
| Case | I | II | III | IV | In EMU-CGS | In SI |
| ΔM | gauss | emu/cm³ | emu/cm³ | emu/mm³ | emu/cm³ | A/m |
| Sample lengths | cm | cm | cm | mm | cm | m |
| Pre-factor | 20 | 20 | 20 A·cm²/emu | 20 | 2 | 2 |
| $J_{c,mag}$ (A/cm²) | $8.83 \cdot 10^8$ | $7.03 \cdot 10^8$ | $7.03 \cdot 10^7$ | $7.03 \cdot 10^5$ <br> or <br> $7.03 \cdot 10^8$ | $7.03 \cdot 10^7$ | $7.03 \cdot 10^7$ |
| Correctness of the result | Wrong | Wrong | Correct value but obtained with a wrong pre-factor | Different values, both wrong, depending on the application of the unit conversion | Correct, independently of the used units | Correct, independently of the used units |

## 4. Conclusions

In summary, the formula for calculation of $J_{c,mag}$ starting from the Magnetization M, within Bean's critical state model, for a rectangular parallelepiped sample with isotropic critical current densities and with volume *w* × *b* × *d*, when the magnetic field is applied perpendicularly to the largest face with dimension *w* by *b* (with *w* ≤ *b*), and with *self-consistent magnetic units* that should be used is:

$$\frac{2 \times \Delta M}{w \times \left(1 - \frac{w}{3b}\right)} = J_{c,mag} \quad (13)$$

where ΔM is the width of the volume magnetization loop. In this equation, the pre-factor, number 2, is *dimensionless* similarly to numbers in any scientific formula or expression. The units of M can be *modified using appropriate magnetic unit conventions* and so can the units of the sample dimensions $w$, $b$, and $d$ to arrive at $J_{c,mag}$ in the desired units, thereby removing any ambiguities related to the units of the pre-factor. Use of this expression will resolve the continued propagation of errors in the literature, is consistent with correct dimensional analysis, and eliminates discrepancies in the calculated $J_{c,mag}$.

**Funding:** M.P. and A. Galluzzi acknowledge support from the EU COST Actions CA19108 Hi- SCALE, CA20116 OPERA, and CA21144 SUPERQUMAP. M.P. acknowledges partial financial support by the PRIN 2022 PNRR Project QUESTIONs Grant No. P2022KWFBH. R. K. and A. Goyal were supported by ONR Grant No. N00014-21-1-2534.